\documentstyle[preprint,aps]{revtex}

\begin{document}

\draft

\title{
Variational state based on the Bethe ansatz solution and \\
a correlated singlet liquid state 
in the one-dimensional $t$-$J$ model
}

\author{
Kenji Kobayashi
}

\address{
Department of Natural Science, Chiba Institute of Technology, \\
      2-1-1, Shibazono, Narashino-shi, Chiba 275, Japan
}

\author{
Chikaomi Ohe and Kaoru Iguchi
}

\address{
Department of Chemistry, School of Science and Engineering, \\
      Waseda University, Tokyo 169, Japan
}

\date{\today}
\maketitle

\begin{abstract}
    The one-dimensional $t$-$J$ model is investigated by
the variational Monte Carlo method.
A variational wave function based on the Bethe ansatz
solution is newly proposed,
where the spin-charge separation is realized,
and a long-range correlation factor of Jastrow-type
is included.
In most regions of the phase diagram, this wave function
provides an excellent description of the ground-state properties
characterized as a Tomonaga-Luttinger liquid;
Both of the amplitude and exponent of correlation functions
are correctly reproduced.
For the spin-gap phase, 
another trial state of correlated singlet pairs with a Jastrow factor 
is introduced.
This wave function shows generalized Luther-Emery liquid behavior,
exhibiting enhanced superconducting correlations 
and exponential decay of the spin correlation function.
Using these two variational wave functions,
the whole phase diagram is determined.
In addition,  relations between the correlation exponent and 
variational parameters in the trial functions are derived.
\end{abstract}


%
%
\section{INTRODUCTION}
    The anomalous properties found in high-$T_c$ superconducting 
copper oxides\cite{Bednorz} have led to a renewal of interest in 
strongly correlated electron systems in low dimensions. 
Among various candidates, the $t$-$J$ model 
has attracted considerable attention as a model to describe 
the cuprate superconductors.\cite{Anderson,ZR}

     For the one-dimensional (1D) $t$-$J$ model, 
much progress has been achieved using analytical and 
numerical techniques.\cite{Review1D}
It has been found\cite{OLSA,Chen93,HM93} that the three main regions 
can be distinguished in the phase diagram defined by the electron 
density  and the ratio of spin exchange interaction to hopping 
amplitude, $J/t$.
First, a Tomonaga-Luttinger liquid\cite{ReviewTLL,TLL} (TLL) holds 
for small $J/t$,
which is characterized by power-law decay of correlation functions.
It has been clarified that the separation of spin and charge degrees of 
freedom is playing an essential role in this region.\cite{OS}
Second, phase separation takes place for large $J/t$,
where the system is separated into  electron-rich and hole-rich phases.
Third, there is a region with a gap in the spin excitation spectrum
for $J/t > 2 $ and at small electron densities.

      On the other hand, in the 2D  $t$-$J$ model,
although some aspects are obtained so far,\cite{Review2D}
many problems are left unresolved.
Particularly, the crucial question is 
whether the features realized in 1D system, 
like the charge-spin separation and/or TLL, 
take place also in 2D system or not.\cite{2DTLL}
To obtain an unified and consistent understanding of the
2D $t$-$J$ model, further progress of research is needed.

       The variational Monte Carlo (VMC) method is one of the most
powerful and transparent approaches to investigate 
strongly correlated electron systems.\cite{YokoReview}
It provides a deeper insight because of its explicit form
of the wave function.
It is very important to construct a better trial function
in the framework of the VMC technique.
One way to obtain further insight into the wave function 
in the 2D $t$-$J$ model is to extend the wave function realized 
in 1D system. 
For this purpose, examining trial wave functions for 1D system 
in detail gives us useful references in the pursuit 
of the 2D $t$-$J$ model.
We shall study variational wave functions in the 1D $t$-$J$ model,
keeping a possibility of extending to the 2D system in mind.

     So far, various kinds of variational functions have been 
proposed for strongly correlated electron 
systems.\cite{VMC,GrosReview,YS90,YO90,HM91,YO96}
The Gutzwiller wave function\cite{Gutz} has been extensively studied 
for its simplicity, and  shown to be a
good trial function for the supersymmetric ($J/t=2$) 
1D  $t$-$J$ model.\cite{YO90}
This wave function was improved for other values of $J/t$
by introducing a conventional Jastrow-correlation factor, 
but the expected TLL behavior was not recovered.\cite{YO90,YO96}
Recently, Hellberg and Mele have introduced a simple trial 
wave function of Jastrow-type.\cite{HM91}
It takes into account an effect of long-range correlations, 
and shows successfully the anomalous power-law behavior 
in correlation functions.
This wave function has been extended
to the 2D $t$-$J$ model to discuss the TLL instability.\cite{VG}
However, the properties of the exact ground-state are not wholly 
reproduced by this wave function, 
especially in the small $J/t$ region.\cite{YO96}

      In this paper we introduce another type of variational wave
functions. Two kinds of trial functions are newly introduced.
First, we consider a variational wave function based on 
the Bethe ansatz solution.
In the limit of $J/t \rightarrow 0$, the charge and spin degrees
of freedom are completely separated, and the ground-state wave
function obtained from the Bethe ansatz can be written 
as a product of the two contributions.\cite{OS}
For finite values of $J/t$, although the charge and spin are separated,
they interact strongly. To take into account this effect,
we introduce a Jastrow-type correlation factor into 
the Bethe ansatz solution for $J/t \rightarrow 0$.
It is shown that this wave function has an advantage of providing 
an excellent description of the ground-state properties 
in most regions of the phase diagram; 
Both of the magnitude and exponent of correlation functions
are correctly reproduced, and a quantitative discussion can be made.

      Next, we consider a trial function for small electron densities.
For the spin-gap phase, Chen and Lee have proposed a variational function
of a gas of noninteracting bound singlet pairs.\cite{Chen93}
This wave function corresponds to a Luther-Emery 
state\cite{LE,ReviewTLL} with infinite correlation exponent.
More accurate trial function can be generated by correlating 
the singlet pairs with a Jastrow factor.
This is just our trial wave function for the spin-gap phase
introduced in this paper.
This wave function shows generalized Luther-Emery liquid behavior,
exhibiting enhanced superconducting correlations 
and exponential decay of the spin correlation function.

      Comparing energies of the trial function based on 
the Bethe ansatz solution and the generalized Luther-Emery state,
the entire phase diagram is determined.
Evaluating the correlation exponents by a finite size scaling analysis,
the relations between the correlation exponent and 
variational parameters in these trial functions are derived.

This paper is organized as follows. In the next section
our trial functions are introduced. Section III provides the 
results of physical quantities by the VMC calculations.
Energies and various correlation functions are compared with the exact
calculations in Sec. III A. 
In Sec. III B, the correlation exponents are evaluated from 
the finite size scaling
to discuss the long-range behavior of correlation functions.
The phase diagram of the 1D $t$-$J$ model 
determined by our wave functions is shown in Sec. III C.
Section VI is devoted to a summary and discussions on
related problems.

\section{TRIAL WAVE FUNCTION}
    The $t$-$J$ model is defined by the Hamiltonian,
\begin{equation}
    H = -t \sum_{\langle ij \rangle \sigma}
           ( \hat{c}_{i \sigma}^\dagger \hat{c}_{j \sigma} 
             + \mbox{\rm H.c.} ) 
       + J \sum_{\langle ij \rangle }
         ( {\bf S}_{i} \cdot {\bf S}_{j} - \frac{1}{4} n_{i}  n_{j} ) , 
      \label{hamil}
\end{equation}
where $ \hat{c}_{i \sigma}^\dagger = 
  c_{i \sigma}^\dagger (1 - n_{i, -\sigma}) $, 
$ c_{i \sigma}^\dagger $ being the creation operator for an electron 
with spin projection $\sigma$ at lattice site $i$, and
 $ n_i=\sum_{\sigma} n_{i \sigma} = 
   \sum_{\sigma} c_{i \sigma}^\dagger c_{i \sigma} $.
Thus $ \hat{c}_{i \sigma}^\dagger $ creates an electron only on 
an empty site, avoiding double occupancy. 
The spin operator associated with site $i$ is defined as
 $ {\bf S}_i = \frac{1}{2} \sum_{\alpha,\beta}
       c_{i \alpha}^\dagger \mbox{\boldmath $\sigma$}_{\alpha,\beta}
       c_{i \beta} $,
where $ \mbox{\boldmath $\sigma$} =( \sigma_x, \sigma_y, \sigma_z)$ 
is a vector of Pauli matrices.
The summations in Eq.~(\ref{hamil}) are taken over nearest 
neighboring pairs.
This model reduces to the  $U=\infty$ Hubbard model
in the limit $J/t \rightarrow 0$.

     For highly correlated electron systems, 
Gutzwiller-Jastrow-type wave functions with two-body correlation 
factor are fairly common.\cite{VMC,GrosReview,YS90,YO90,HM91,YO96}
The Gutzwiller wave function,\cite{Gutz}
 which is a prototype of the trial function of this type, 
 is often used 
as a starting trial function for its simplicity. It is defined as 
\begin{equation}
  \left| \psi_{{\rm G}} \right\rangle =
         P_d \left| \phi_{{\rm F}} \right\rangle 
        =  \prod_{i} \left( 1- n_{i \uparrow} n_{i \downarrow} \right) 
           \left| \phi_{{\rm F}} \right\rangle   ,    \label{psiG}
\end{equation}
where $\phi_{{\rm F}}$ is a simple Fermi sea 
and $P_d$ is the operator projecting out the double occupancy.
This wave function is essentially a Fermi-liquid state,
having a discontinuity in momentum distribution at $k=k_F$.
Thus, the expected TLL behavior was not recovered.\cite{YO90}
Hellberg and Mele have introduced a variational state
with a long-range correlation:\cite{HM91}
\begin{equation}
  \left| \psi_{{\rm HM}} \right\rangle =
           \prod_{i \ne j} \prod_{\sigma \sigma '} \;
           \bigl[ 1- \left( 1-|d_{ij}|^{\nu} \right)
                     n_{i \sigma} n_{j \sigma '} \bigr]
           \left| \psi_{{\rm G}} \right\rangle   ,    \label{psiHM}
\end{equation}
\begin{equation}
    d_{ij} = \sin \left[ \pi r_{ij} / N_s \right] 
                                             ,    \label{dij}
\end{equation}
where  $r_{ij}=|r_i-r_j|$  is the distance between 
the $i$-th and $j$-th sites, and $N_s$ the number of sites.
When $\nu=0$, $\psi_{{\rm HM}}$ reduces to $\psi_{{\rm G}}$.
It has been shown that $\psi_{{\rm HM}}$ exhibits the characteristic 
behavior of TLL.\cite{HM91}
However, the correlation exponent estimated with this wave function
does not coincide with the exact value for small $J/t$.\cite{YO96}
This disagreement becomes apparent 
when the global features of various correlation functions are
compared with the exact ones.\cite{YO96}

       An important feature of the TLL is the separation of spin 
and charge degrees of freedom in the low-energy excitations. 
In the limit of $J/t \rightarrow 0$, Ogata and Shiba have shown that 
the ground-state wave function obtained from the Bethe ansatz
has a simple form due to the complete decoupling of
 charge and spin degrees of freedom.\cite{OS}
It can be written as a product of a Slater determinant of spinless
fermions describing the charge degrees of freedom and the spin wave
function of the squeezed Heisenberg model in which all empty sites are
omitted.
The ground-state wave function in the limit of $J/t \rightarrow 0$
is expressed as
\begin{equation}
   \psi_0 ( x_1,\dots, x_{N_e} )
     = X( x_1,\dots, x_{N_e}) Y( y_1,\dots, y_{M} )  ,    \label{psi0}
\end{equation}
where
\begin{equation}
    X( x_1,\dots, x_{N_e} ) = \det \left[ \exp ( i q_i x _j ) \right] ,
         \label{X}
\end{equation}
$\left\{ x_j \right\}$ are the positions of $N_e$ electrons, 
and $\left\{ y_j \right\}$ are the coordinates of $M$ up spins 
with vacant sites  omitted.
$X( x_1,\dots, x_{N_e} )$ is the wave function for
noninteracting spinless fermions with momenta $\left\{ q_i \right\}$,
and $Y( y_1,\dots, y_{M} )$ is the ground-state wave function of
 the $S=\frac{1}{2}$ antiferromagnetic Heisenberg model.

    For finite values of $J/t$, the charge and spin degrees of freedom 
are no longer completely separated, and the charge-spin coupling occurs.
Thus we introduce a Jastrow-type correlation factor in our trial
wave function to mix the charge and spin degrees of freedom.
We shall study the following variational state for the 
1D $t$-$J$ model:
\begin{equation}
  \left| \psi_{{\rm BA}} \right\rangle =
           F_J  \left| X Y \right\rangle   ,    \label{psiBA}
\end{equation}
where the amplitude of 
 $ \left| X \right\rangle = 
 \prod_i ^{N_e} c_{q_i}^\dagger  \left| 0 \right\rangle $
 is given by Eq.~(\ref{X}),
ensuring the absence of double occupancy.
The long-range correlation factor of Jastrow-type
in Eq.~(\ref{psiBA}) is defined as
\begin{equation}
  F_J = \prod_{i \ne j} \prod_{\sigma \sigma '} \;
       \bigl\{ 1- \left[ 1-\eta (r_{ij} ; \sigma \sigma ') \right]
                     n_{i \sigma} n_{j \sigma '} \bigr\}  , 
      \label{FJ}
\end{equation}
and the form of function $\eta$ is assumed to be
\begin{equation}
   \eta ( r_{ij} ; \sigma  \sigma' )= \left\{
        \begin{array}{ll}
            |d_{ij}|^{\nu_1} \; , & \quad \mbox{if}  \; 
                               \sigma  =    \sigma'  \\
            |d_{ij}|^{\nu_2} \; , & \quad \mbox{if}  \; 
                               \sigma  \ne  \sigma'  
        \end{array} \right.  \; ,       \label{eta}
\end{equation}
where $ d_{ij} $ is given by Eq.~(\ref{dij}).
The Jastrow factor $F_J$ modulates the Bethe ansatz
wave function by the distance between all pairs of particles.
Positive value of $\nu_1$ ($\nu_2$) induces a repulsive correlation 
between particles with the same spins (opposite spins),
while the negative values provide an attractive correlation.
When $\nu_1=\nu_2$, $F_J$ reduces to the correlation factor
studied by Hellberg and Mele.

     The wave function of the spin part in Eq.~(\ref{psiBA})
is approximated as a trial function of 
Jastrow-Marshall-type,\cite{Marshall}
\begin{equation}
    Y( y_1,\dots, y_{M}) 
     =  (-1)^{ L \left\{ y_i \right\} }
         \prod_{i<j} |s_{ij}|^{\nu_s}   ,  \label{Y}
\end{equation}
where $ s_{ij} = \sin \left[ \pi ( y_i-y_j) / N_e \right] $ 
for a system of $N_e$ electrons and $ L \left\{ y_i \right\}$ is
the number of up spins in one sublattice contained in the spin
configuration $( y_1,\dots, y_{M})$.
With this trial function $Y( y_1,\dots, y_{M})$, 
we have calculated the ground-state energies of 
the 1D antiferromagnetic Heisenberg model,
 $  H_{{\rm Heis}} =  J \sum_{\langle ij \rangle }
            {\bf S}_{i} \cdot {\bf S}_{j}  $ 
with $N_e \le 70$ by the VMC technique,
and estimated the energy in the thermodynamic limit from 
finite size scaling with a formula
$  E/N_e   =  E_{\infty} + C/N_e^2 $.
The minimum energy is realized for $\nu_s \approx 2$,
and the resultant energy is 
$ E_{\infty} = (-0.4421 \pm 0.0001) J $,
which is quite close to the exact value 
by the Bethe ansatz,\cite{Bethe}
$ E_{{\rm BA}}/N_e 
  = -(\ln 2 -\frac{1}{4} ) J = -0.443147\dots J $.
The difference is only 0.24 \%.
Therefore, $Y( y_1,\dots, y_{M})$ well reproduces the true
ground-state wave function of the 1D Heisenberg model.

      As a result, we have three variational parameters in our trial
state (\ref{psiBA}), i.e., $\nu_1$, $\nu_2$, and $\nu_s$. 
In most regions of the phase diagram, this wave function
$\psi_{{\rm BA}}$ successfully reproduces the exact ground state
of the 1D $t$-$J$ model as shown in Sec. III.

      However,  a Luther-Emery liquid behavior,
 exhibiting a gap in the spin excitation spectrum and enhanced
superconducting correlations, is found for $J > 2t$ and
at small densities.\cite{Chen93,HM93}
The true ground state for this region lies out of the variational
subspace spanned by $\psi_{{\rm BA}}$.
To represent the spin-gap phase better, we introduce another
trial state as follows.
\begin{equation}
  \left| \psi_{{\rm RVB}} \right\rangle =  F_J P_d
            \sum_{ \{ i_n j_n \} } \prod ^{N_e/2}_n h^{r_{i_n j_n}-1}
                 [i_n,j_n] 
             \left| 0 \right\rangle 
                     ,    \label{psiRVB}
\end{equation}
where
 $  [i,j]  =
         (   c_{i \uparrow}^{\dagger} c_{j \downarrow}^{\dagger}
           - c_{i \downarrow}^{\dagger} c_{j \uparrow}^{\dagger} ) 
               $
is a singlet pair in a given configuration $\{ i_n j_n \}$,
and $P_d$ projects out the double occupancy.
In Eq.~(\ref{psiRVB}),
$ h^{r_{i j}-1} $ controls a weight for a singlet bond 
as a function of its length.
The function $\eta$ in the Jastrow factor is taken to be
$ \eta ( r_{ij} ; \sigma  \sigma' )= |d_{ij}|^{\lambda} $,
 i.e., $F_J$ is assumed to be spin-independent.
 Two variational parameters,  $\lambda$ and $h$, 
are contained in the trial function $\psi_{{\rm RVB}}$.
This is a natural generalization of the wave function
of a gas of noninteracting bound singlet pairs 
proposed by Chen and Lee,\cite{Chen93} which
corresponds to a Luther-Emery state with infinite 
correlation exponent.
Correlating the singlet pairs with the Jastrow factor $F_J$,
$\psi_{{\rm RVB}}$ can be expected to exhibit generalized 
Luther-Emery behavior.
It is also a particular form of the 
resonating valence bond (RVB) state.
In fact, $\psi_{{\rm RVB}}$ can be rewritten as
\begin{equation}
  \left| \psi_{{\rm RVB}} \right\rangle =  F_J P_d
         \left[   \sum_k \frac{\cos k - h}{h^2-2 h \cos k +1}
            c_{k \uparrow}^{\dagger} c_{-k \downarrow}^{\dagger}
             \right]^{N_e/2} 
             \left| 0 \right\rangle 
                     ,  
\end{equation}
which explicitly shows a singlet liquid picture of 
the RVB state.\cite{Anderson}

\section{SIMULATION AND RESULTS}
     In this section, we present the results of the VMC calculations 
for various values of $J/t$ and $n_e=N_e/N_s$ with $N_e$ and $N_s$
being the number of electrons and sites, respectively,
and make comparisons with
those of exact calculations and other trial functions.
We consider the 1D $t$-$J$ model with up to 300 sites
under the periodic boundary condition with $N_e/2=\mbox{odd}$.

     Variational parameters in Eqs.~(\ref{psiBA}) and (\ref{psiRVB})
are optimized using a conjugate-gradient method
combined with the fixed sampling in the VMC calculations.
Technical details of the optimization procedure were described 
in Ref.~\onlinecite{Koba93}, and some practical improvements
are made to achieve the convergence rapid enough to handle 
 multiparameter optimization: a quasi-Newton algorithm is 
employed instead of Powell's optimization algorithm
in Ref. \onlinecite{Koba93}, 
and the gradient is evaluated by the numerical differentiation.
Once the fully optimized wave function is obtained, 
we use it in evaluating the physical quantities with another VMC run
in order to examine the properties of the 1D $t$-$J$ model in detail.
Calculated quantities are the total energy per site,
the momentum distribution function,
\begin{equation}
    n(k) = \frac{1}{2N_s} \sum_{ij \sigma} e^{ik(r_i-r_j)}
           \left\langle c_{i \sigma}^\dagger c_{j \sigma} \right\rangle 
                     ,    \label{nk}
\end{equation}
and the equal-time correlation functions,
where $ \langle \cdots \rangle $ indicates the expectation value
for a given trial function.
The spin- and charge- correlation functions
in Fourier-transformed form are defined as
\begin{equation}
    S(k) = \frac{4}{N_s} \sum_{ij} e^{ik(r_i-r_j)}
           \left\langle S_i^z S_j^z \right\rangle 
                     ,    \label{Sk}
\end{equation}
\begin{equation}
    C(k) = \frac{1}{N_s} \sum_{ij} e^{ik(r_i-r_j)} \Bigl[
           \left\langle n_{i} n_{j} \right\rangle
         - \langle n_{i} \rangle  \langle n_{j} \rangle
                                                     \Bigr]
                     ,    \label{Ck}
\end{equation}
respectively.
The singlet pairing correlation function is defined as
\begin{equation}
    P(k) = \frac{1}{N_s} \sum_{ij} e^{ik(r_i-r_j)} 
           \left\langle \Delta_{i}^\dagger
                        \Delta_{j}   \right\rangle   
                     ,    \label{Pk}
\end{equation}
where $\Delta_{i}$ is the annihilation operator of 
a nearest neighboring electron singlet pair,
\begin{equation}
     \Delta_{i} = \frac{1}{\sqrt{2}} 
         (   c_{i \uparrow} c_{i+ 1 \downarrow}
           - c_{i \downarrow} c_{i+ 1 \uparrow} ) 
                     .    \label{Di}
\end{equation}
We collect typically 30,000 samples to take averages of the
energy for the optimization of variational parameters,
and 100,000 -- 200,000 samples for the evaluations 
of the expectation values of observables.

\subsection{Quarter-filled case}
    First we compare the properties of $\psi_{{\rm BA}}$ 
with those of $\psi_{{\rm HM}}$ 
for the quarter-filled case: $n_e=\frac{1}{2}$.
At this electron density, the spin-gap state is absent.
In addition, when $ J/t \rightarrow 0$, the exact results of
correlation functions have been obtained\cite{OS,YO90,PS} for 
fairly large $N_s$, with which we can compare the VMC results. 

    The result of the optimization of variational parameters 
in $\psi_{{\rm BA}}$ is shown in Fig.~\ref{figopt} for $n_e=1/2$ 
as a typical case.
The data in the region $J/t \le 3.3$ are fitted to polynomials 
of degree $m$, where $m=2$ for $\nu_2$ and $\nu_s$,
and $m=3$ for $\nu_1$, respectively.
For $ J/t \rightarrow 0$, the minimum energy is realized 
for $\nu_1=\nu_2=0$ and $\nu_s \approx 2$.
In the case of finite $J/t$, the optimal variational parameters 
show the coupling of charge and spin degrees of freedom
as expected, i.e., $\nu_1 \neq 0$ and/or $\nu_2 \neq 0$.
$\nu_2$ and $\nu_s$ decrease
with $J/t$ while the dependence of $\nu_1$ on $J/t$ is week 
for $J/t \le 2$.
Near $J/t=2$, $\nu_s$ intersects the zero line, 
and $\nu_2$ becomes $-1$.
For larger $J/t$, the attractive correlation between electrons 
with opposite spins is prominent.
For $J/t > 3.3$, the variational state separates into 
the electron rich and poor phases.
The variational parameters abruptly change their behavior at the
phase separation boundary as shown in Fig.~\ref{figopt}.
The data in the phase separated region are fitted 
to other polynomials.

      For other values of $n_e$, the optimal values of 
variational parameters show similar behaviors
except for curvatures.
Notice that the trial state with optimal variational parameters 
is singlet although the Jastrow factor $F_J$ is spin-dependent;
The VMC evaluation of the physical quantities 
shows that the total spin is zero, and the spin correlation 
function is isotropic ($S^{xx}(k)=S^{yy}(k)=S^{zz}(k)$),
as far as the optimized $\psi_{{\rm BA}}$ is employed.\cite{Note2}

      Next we discuss the variational energies.
It has been shown that the ground-state energy converges 
smoothly to the thermodynamic limit.\cite{YO90,YO96} 
Following Yokoyama and Ogata,\cite{YO90,YO96} 
we estimate the variational energy per site in the 
limit $N_s \rightarrow \infty$ from the finite size scaling.
We calculate the variational energies of 12-, 20-, 36-, 60-, and 100- 
site systems for $n_e=0.5$ using the optimized $\psi_{{\rm BA}}$ and
$\psi_{{\rm HM}}$, and then fit the results to the formula
\begin{equation}
   E/N_s  =  E_{\infty} + C_1/N_s^2  + C_2/N_s^4  + C_3/N_s^6 
                     .    \label{fite}
\end{equation}
The fitted values of $E_{\infty}(\psi_{{\rm BA}})$ and  
$E_{\infty}(\psi_{{\rm HM}})$ are listed in Table~\ref{tableI} 
for several values of $J/t$.
They are compared to the exact results obtained from 
the Bethe ansatz for $J/t=0$\cite{Shiba} and 2,\cite{J2}
and the extrapolated values of the exact diagonalization of small 
clusters for $J/t=1$ and 3. The latter is evaluated from fitting 
the energies of 4-, 8-, 12-, and 16- site clusters 
to Eq.~(\ref{fite}).\cite{YO96}

     Using $\psi_{{\rm BA}}$, the ground-state energy per site 
in the limit  $N_s \rightarrow \infty$ is obtained to be 
$E_{\infty}= -2t/\pi $ for $J/t \rightarrow 0$,
equivalent to the exact energy.
The reason for the coincidence is that the energy is determined only by
the charge degree of freedom in the limit $ J/t \rightarrow 0$,
whose treatment is rigorous in $\psi_{{\rm BA}}$.
In fact, the variance 
$ \langle \left( H - E(\psi_{{\rm BA}}) \right)^2 \rangle $ 
in VMC sweeps vanishes at any $n_e$.
For all the range of $J/t$, $E_{\infty} (\psi_{{\rm BA}})$ is 
quite close to the exact energy, as shown in Table~\ref{tableI}.
 Especially in the small $J/t$ region, the advantage of 
$\psi_{{\rm BA}}$ over $\psi_{{\rm HM}}$ is obvious.
The difference in energy between $\psi_{{\rm HM}}$ and 
the exact one is largest for $J/t = 0$, while
the error of $\psi_{{\rm BA}}$ gradually increases with $J/t$ 
except for $J/t=2$.

    For $J/t=2$, both of $E_{\infty} (\psi_{{\rm BA}})$ and 
$E_{\infty} (\psi_{{\rm HM}})$ are extremely close to 
the exact energy obtained by the Bethe ansatz.
In this connection, Yokoyama and Ogata have shown 
that the Gutzwiller function $\psi_{{\rm G}}$ is a
good trial function for $J/t=2$ and all range of $n_e$.\cite{YO90}
In fact, the analytical calculation of the energy using the 
Gutzwiller wave function shows 
$E_{\infty} (\psi_{{\rm G}})=-0.903092\dots$,\cite{MV,YO90}
which is close to the exact one.
Moreover, Kuramoto and Yokoyama\cite{KY} have found 
that $\psi_{{\rm G}}$ is the exact ground state of 
the $t$-$J$ model with long-range interactions
and hoppings satisfying $ J_{ij} = 2 t_{ij} \propto r_{ij}^{-2}$.
In the supersymmetric case, the ground state behaves as almost 
a free-electron state\cite{YO96} except for the exclusion 
of double occupancy because of the cancellation of hopping and 
interacting processes, both of which have the same weight $t=J/2$.
Therefore, it is natural to expect that variational energies
converge to the same value
if a trial function can recover the ``free-electron" nature for $J/t=2$.
The above three variational wave functions have this property correctly
although the long-range behavior of correlation functions is different.

     In Fig.~\ref{figcf0}, (a) the momentum distribution function $n(k)$,
(b) the spin correlation function $S(k)$,
(c) the charge correlation function $C(k)$, and
(d) the singlet pairing correlation function $P(k)$ 
are shown for $ J/t = 0$, 
where open circles denote the VMC results of $\psi_{{\rm BA}}$,
and the broken lines represent those of $\psi_{{\rm HM}}$.
We evaluate the data for the lattice with 100 sites 
at quarter filling.
The exact results obtained from the Bethe ansatz solution 
for $N_s=52$ are also shown in Fig.~\ref{figcf0} 
with solid lines.\cite{OS,YO90,PS}
In the TLL, the momentum distribution function exhibits 
 power-law singularities at $k_F$ and $3k_F$ 
although the latter is very weak.
Qualitatively, both of $\psi_{{\rm BA}}$ and $\psi_{{\rm HM}}$ reproduce
the anomalous power-law behavior  inherent in
TLL as shown in Fig.~\ref{figcf0} (a).
However, $n(k)$ by $\psi_{{\rm HM}}$ departs appreciably from
that by the Bethe ansatz while $n(k)$ by $\psi_{{\rm BA}}$ 
almost coincides with the exact one.
For the correlation functions, there are also the apparent differences
between $\psi_{{\rm HM}}$ and the exact result while $\psi_{{\rm BA}}$ is
quite close to the exact one, as shown in Figs.~\ref{figcf0} (b)--(d).
In particular, $C(k)$ calculated with  $\psi_{{\rm HM}}$ 
shows different behavior: it has a small cusp at $2k_F$.
On the other hand, both of $\psi_{{\rm BA}}$ and the exact result
exhibit linear $k$-dependence in $C(k)$ 
reflecting the free spinless fermions nature.
Thus $\psi_{{\rm BA}}$ recovers the global features of correlation 
functions correctly in contrast to $\psi_{{\rm HM}}$.

    In Fig.~\ref{figcf123}, we show the VMC results of 
the momentum distribution function and the correlation functions
for finite $J/t$ and $n_e=\frac{1}{2}$.
We have plotted the results of $\psi_{{\rm BA}}$ (open symbols) 
and $\psi_{{\rm HM}}$ (broken lines) for comparison.
At $J/t=1$(open diamonds), the difference between $\psi_{{\rm BA}}$ 
and $\psi_{{\rm HM}}$ is observed, but it is not so large as 
in the case of $J/t = 0$ except for the peak of $C(2k_F)$.
For $J/t=2$(open squares), the two variational functions 
give almost the same results,
in accordance with the agreement of energies. 
At $J/t=3$(open triangles), 
it is remarkable that $P(k)$ at $k=0$ is strongly enhanced 
due to the increased effective attraction between neighboring
electrons with opposite spins as shown in Fig.~\ref{figcf123} (d).
One can also see the  enhancement of $P(k)$ near $k=\pi$ 
as $J/t$ increases.
The singularity of $P(2k_F)$ becomes sharp when $J/t=2$.

The correlation functions by $\psi_{{\rm BA}}$
shown in Fig.~\ref{figcf123} 
agree well with the result of the Quantum Monte Carlo 
calculation\cite{AW} or of the exact diagonalization 
in smaller lattices.\cite{YO90,YO96,PS}

\subsection{Correlation exponent}
Let us now examine the correlation exponents
to discuss the long-range behavior of correlation functions.
In the TLL regime, both the charge and spin excitations 
are gapless and the correlation functions show power-law decay.
Following the TLL theory,
the leading singularities of the distribution function and 
correlation functions can be 
written as follows.\cite{Review1D,ReviewTLL}
\begin{eqnarray}
   n(k) &\sim& |k-k_F|^{\alpha} \mbox{sgn} (k-k_F) 
                                               \quad \mbox{for}  \;
           k \sim  k_F   \; ,  \\
   S(k) &\sim& |k-2k_F|^{\eta-1}               \quad \mbox{for}  \;
           k \sim 2k_F   \; , \label{expSk} \\
   C(k) &\sim& |k-2k_F|^{\eta-1}               \quad \mbox{for}  \;
           k \sim 2k_F   \; , \\
   P(k) &\sim& |k|^{\mu-1}                     \quad \mbox{for}  \;
           k \sim 0      \; .
\end{eqnarray}
Logarithmic corrections have been omitted in these formulae.
As far as the interaction is isotropic in spin space,
the critical exponents are described by a dimensionless 
TLL parameter $K_{\rho}$ as follows.\cite{Review1D,ReviewTLL}
\begin{eqnarray}
   \alpha &=& \min \left[ 1, (K_{\rho}-1)^2/(4K_{\rho}) \right]   ,  \\
   \eta   &=& K_{\rho} + 1                     ,  \label{expeta}     \\
   \mu    &=& 1/K_{\rho} + 1                   .  \label{mu}
\end{eqnarray}

   The correlation exponent $K_{\rho}$ can now be rather easily 
calculated from our VMC result of $P(k)$ than $n(k)$, $C(k)$, or $S(k)$
since the singularities of these quantities
become much weaker than that of $P(k)$ in some cases.
Following Assaad and W\"urtz,\cite{AW}
we use the following procedure to obtain the exponent of $P(k)$.
Let us assume a behavior
\begin{equation}
    \ln |P(0) - P(k_1)| = - (\mu-1) \ln N_s + a    ,  \label{fitP}
\end{equation}
where $k_1=2 \pi /N_s$, and $\mu$ and $a$ are the fitting parameters.
Fitting the data for various lattice sizes to Eq.~(\ref{fitP}),
the exponent of $P(k)$ can be evaluated
by finite size scaling.
Once $\mu$ is obtained, $K_{\rho}$ is given by Eq.~(\ref{mu}).

As an  example, 
we plot the fitting result of $P(k)$ for $n_e=\frac{1}{2}$ 
in Fig.~\ref{figfp}.
We have calculated $P(k)$ of 12-, 20-, 36-, 60-, and 100-
site systems for $n_e=\frac{1}{2}$ 
using the optimized $\psi_{{\rm BA}}$, 
and then fit the results to Eq.~(\ref{fitP}).
The linearity of these plots is good.
The correlation exponent $K_{\rho}$ obtained from the slope of 
these plots and Eq.~(\ref{mu}) is tabulated in Table~\ref{tableII}.
For $\psi_{{\rm HM}}$, one can rather analyze the exponent directly.
The relation between the variational parameter in $\psi_{{\rm HM}}$
and $K_{\rho}$ has been derived analytically from the generalized 
conformal field theoretical argument to be\cite{expHM}
\begin{equation}
   K_{\rho} = \frac{1}{2\nu+1}     .       \label{Khm}
\end{equation}
Therefore, the optimal value of $\nu$ can be used to determine 
$K_{\rho}$ for $\psi_{{\rm HM}}$.
The results for $\psi_{{\rm HM}}$ in Table~\ref{tableII}
are obtained from this equation.
The expected values of the exponents are also shown 
in Table~\ref{tableII}: 
for $J/t=0$ and $2$, $K_{\rho}$ has been exactly determined from 
the bosonization theory\cite{Schulz} or 
the conformal field theory,\cite{CFTJ0,CFTJ2} with
$K_{\rho}=0.5$ and $0.85$, respectively,
and the exact diagonalization on a 16-site ring has shown\cite{OLSA}
that $K_{\rho} \approx 0.62$ and $2.2$ for $J/t=1$ and 3, respectively.

The results of $\psi_{{\rm HM}}$ and $\psi_{{\rm BA}}$
are consistent with the expected values;
At $J/t=3$, $K_{\rho}$ becomes larger than 1 and thus the superconducting
correlations correspondingly dominate the long-range
behavior while for smaller value of $J/t$, $K_{\rho} <1$.
However, the quantitative coincidence is not so good
for $\psi_{{\rm HM}}$,
while the exponents evaluated from $\psi_{{\rm BA}}$ 
are surprisingly close to the expected values.
Therefore, one could conclude that
our variational wave function $\psi_{{\rm BA}}$ can quantitatively
reproduce not only the amplitude of correlation functions,
but also the correlation exponent.
The last column in Table~\ref{tableII} shows the value of 
$(\nu_1+\nu_2 + 2)^{-1}$ evaluated from the optimized 
variational parameters in $\psi_{{\rm BA}}$.
It seems that $ K_{\rho}$ agrees with $(\nu_1+\nu_2 + 2)^{-1}$.
This point will be discussed in the next subsection.

\subsection{Phase diagram}
The phase diagram of the 1D $t$-$J$ model obtained by
$\psi_{{\rm BA}}$ and $\psi_{{\rm RVB}}$ is shown in Fig.~\ref{figpd}
with contour lines for several values of $K_{\rho}$.
As seen in Fig.~\ref{figpd}, there are four distinct phases.
For small $J/t$, the ground state is a repulsive TLL
with $K_{\rho} < 1$.
In this region, spin correlations dominate the long-range behavior.
Increasing $J/t$,  these correlations are suppressed, 
and the ground state changes to an attractive TLL
with $K_{\rho} > 1$. It has dominant singlet pairing correlations.
For larger $J/t$, the variational state is phase-separated
into the electron-rich phase with antiferromagnetic order
and the empty phase.
In this region, the longest wavelength charge correlation 
$C(k_1=2\pi/N_s)$ diverges when the system size $N_s$ is increased. 
This behavior is in contrast with the TLL, 
where $C(k_1)$ remains finite.
The phase separation boundary in Fig.~\ref{figpd} is determined
by the behavior of $C(k_1)$.
These three phases are described by $\psi_{{\rm BA}}$.
The spin-gap state lies between the TLL and 
phase separated regions at small densities,
where $\psi_{{\rm RVB}}$ is more stable than $\psi_{{\rm BA}}$.
The optimal values of variational parameters in $\psi_{{\rm RVB}}$ 
fit to the formulae as $h=1.037-0.151(J/t)$ and 
$\lambda=0.781-0.173(J/t)$ for $n_e=0.2$ and $2.8 \le J/t \le 3.2$,
for example.

The phase diagram determined by the present method
is consistent with the result of the exact diagonalization 
on a 16-site ring.\cite{OLSA}
It is correctly recovered that $K_{\rho}=1/2$ at any electron density
in the limit  $ J/t \rightarrow 0$.\cite{Schulz,CFTJ0}
When we compare the spin-gap region with the result of the wave 
function of noninteracting singlet pairs by Chen and Lee,\cite{Chen93}
the correlation effect of the Jastrow factor introduced 
in $\psi_{{\rm RVB}}$ seems to play no essential role since these 
two trial functions give almost the same result;
The correlation effect only slightly pushes the spin-gap region 
to larger values of $n_e$.
This is because the balance of energy hardly changes, i.e.,
the magnitude of correlation energy induced by the Jastrow factor
in $\psi_{{\rm RVB}}$ is less than 2\%,
which is comparable to the energy lowering of $\psi_{{\rm BA}}$
compared to  $\psi_{{\rm HM}}$.
However, there is a significant difference 
in the long-range behavior of correlation functions.
The wave function by Chen and Lee 
gives  $K_\rho = \infty$ while  $K_\rho $ is finite 
for our $\psi_{{\rm RVB}}$ as seen below.

Figure~\ref{figcf02} shows the distribution function and 
correlation functions for 150 sites and 30 electrons. 
The selected values of $J/t$ are 0, 2.5, and 3
as typical cases of the repulsive TLL, attractive TLL, and
spin-gap state, respectively.
Triangles and squares in Fig.~\ref{figcf02} are evaluated with 
$\psi_{{\rm BA}}$ while circles with $\psi_{{\rm RVB}}$.
The spin correlation functions for $J/t=0$ and $2.5$
exhibit the linear behavior at small $k$ characteristic
of TLL as shown in Fig.~\ref{figcf02} (b).
On the other hand, $S(k)$ for $J/t=3$ is quadratic at small $k$
and analytic for all wave vectors.
Unlike TLLs, a Luther-Emery liquid exhibits
exponential decay of the spin correlation function in real space,
while both charge and singlet pairing correlations decay 
algebraically.\cite{ReviewTLL}
The short-range behavior of $S(k)$ for $J/t=3$ 
shows that the spin-gap state is characterized 
as a Luther-Emery liquid.
The charge correlation function for $J/t=3$ exhibits 
a cusp at $k=2k_F$, indicating the formation of 
bound singlet pairs.
As $J/t$ decreases, this cusp is suppressed.
More definitive characters of the three phases can be
seen in the singlet pairing correlation functions plotted 
in Fig.~\ref{figcf02} (d).
$P(k)$ is fully suppressed when $J/t=0$.
As $J/t$ increases, $P(k=0)$ becomes much larger,
indicating the growth of long-range order.
The cusp at $k=0$ is greatly enhanced for $J/t=3$.

Finally, we plot the exponents of 
the singlet pairing correlation function for $n_e=0.2$ 
as a function of $J/t$ in Fig.~\ref{figexp},
where $\mu-1$ is evaluated from fitting $P(k)$ to Eq.~(\ref{fitP})
for lattice sizes ranging from $N_s=30$ to 150.
In the region $2.8 \le J/t \le 3.2$, the variational energy 
of $\psi_{{\rm RVB}}$ is lower than that of $\psi_{{\rm BA}}$.
Correspondingly, squares and circles in Fig.~\ref{figexp} are evaluated with 
$\psi_{{\rm BA}}$ and $\psi_{{\rm RVB}}$, respectively.
In addition,  $\nu_1+\nu_2+2$ for $\psi_{{\rm BA}}$
and $ 2 \lambda -1$ for $\psi_{{\rm RVB}}$ are shown 
in Fig.~\ref{figexp} with solid lines, 
where $\nu_1$,  $\nu_2$, and  $\lambda$ are optimized variational parameters.
It seems that the fits of $\nu_1+\nu_2+2$ and $ 2 \lambda -1$
to the exponents of $P(k)$ are good.
The exponents of the correlation functions in the Luther-Emery liquid
that decay with power laws can be also described
by a single parameter $K_{\rho}$ like TLL,\cite{ReviewTLL}
and it holds that
\begin{equation}
   \mu    = \frac{1}{K_{\rho}}       .    \label{muLE}
\end{equation}
Using Eqs.~(\ref{mu}) and (\ref{muLE}),
one can conclude from Fig.~\ref{figexp} 
that $K_{\rho}$ relates to the variational parameters 
in $\psi_{{\rm BA}}$ and $\psi_{{\rm RVB}}$ as
\begin{equation}
   K_{\rho} = \frac{1}{\nu_1+\nu_2 + 2} 
         \qquad \mbox{for}  \;  \psi_{{\rm BA}}
                                 ,       \label{Kba}
\end{equation}
and
\begin{equation}
   K_{\rho} = \frac{1}{2 \lambda } 
         \qquad \mbox{for}  \;  \psi_{{\rm RVB}}
                                 ,       \label{Krvb}
\end{equation}
respectively.
The critical exponent $K_{\rho}$ evaluated from $P(k)$ for $n_e=0.2$
is tabulated in Table~\ref{tableIII} together with the predicted 
values by Eqs.~(\ref{Kba}) and (\ref{Krvb}), and they agree very well.
For other values of $n_e$, $K_{\rho}$ also agrees with 
$(\nu_1+\nu_2 + 2)^{-1}$ or $ (2 \lambda )^{-1}$.
The example for $n_e=0.5$ is shown in Table~\ref{tableII}.
$K_{\rho}$ shown in Fig.~\ref{figpd} is evaluated from  Eq.~(\ref{Kba}).

The relations (\ref{Kba}) and (\ref{Krvb}) are confirmed
by the following facts.
(i) In the limit $n_e \rightarrow 1$, we have $\nu_1=\nu_2=0$ and 
$\nu_s \approx 2$, i.e., $\psi_{{\rm BA}}=Y$ for the Heisenberg chain
since $X$ becomes only a constant.
This is in accordance with $ K_{\rho} \rightarrow 1/2$ as 
$ n_e \rightarrow 1$.\cite{OLSA}
(ii) It is correctly recovered that $K_{\rho} \rightarrow 1/2$ 
at any electron density in the limit  
$ J/t \rightarrow 0$,\cite{Schulz,CFTJ0} where we have $\nu_1=\nu_2=0$.
(iii) The magnitude of the discontinuous jump in $\mu -1$ 
at $J/t \sim 2.8$ is $(\nu_1+\nu_2+2)-(2 \lambda -1) \approx 1$ 
as seen in Fig.~\ref{figexp}. This is consistent with the crossover from 
the TLL to the Luther-Emery liquid behavior
described by Eqs.~(\ref{Kba}) and (\ref{Krvb}).
(iv) $\nu_1+\nu_2+2 \approx 0$ on the phase separation boundary
in Fig.~\ref{figpd}, which leads to $K_{\rho}=\infty$.
This corresponds to the divergence of $C(k_1)$.

\section{SUMMARY AND DISCUSSION}
In this paper, we have carried out the VMC calculation 
for the 1D $t$-$J$ model.
As a trial state, we have proposed a new type of
variational wave function based on the Bethe ansatz solution.
In this wave function, the separation of charge 
and spin degrees of freedom is realized explicitly,
and the long-range correlation factor of Jastrow-type
is included.
With this wave function,
it has been shown that the remarkable improvement is achieved 
especially in the small $J/t$ region: 
the variational energy, momentum distribution function, 
and various correlation functions exhibit an excellent coincidence with 
exact ones.
The evaluation of correlation exponents with the finite size scaling
has shown that this variational wave function can correctly reproduce 
not only the global features of correlation functions
but also the long-range behavior with anomalous power-law decay,
which is characteristic of TLL.

In addition, a variational wave function of singlet pairs correlated 
with a Jastrow factor has been introduced to describe the spin-gap phase.
This wave function correctly exhibits enhanced superconducting 
correlations and exponential decay of the spin correlation function,
as expected for the generalized Luther-Emery state.

      Comparing the energies of the trial function based on 
the Bethe ansatz solution and the generalized Luther-Emery state,
the whole phase diagram has been determined.
The VMC results show that our wave functions provide a more precise 
description of the ground-state properties for the 1D $t$-$J$ model 
in the whole phase diagram.
Evaluating the correlation exponents by the finite size scaling 
analysis, the relations between the exponent $K_{\rho}$ and 
the variational parameters in the trial functions have been 
established.

Let us now compare our trial wave functions with others.
For strongly correlated electron systems, the Gutzwiller-Jastrow-type
trial state has been extensively studied, but the conventional
Jastrow factor does not recover the expected TLL behavior
if only short-range correlations are included.\cite{YO90,YO96}
A trial wave function of this type is essentially a Fermi liquid state.
The wave function introduced by Hellberg and Mele successfully 
exhibits the power-law singularity of TLL.\cite{HM91}
The long-range nature of the Jastrow factor is
essential for the non-Fermi liquid behavior.
However, the correlation exponent does not coincide with the exact value.
This disagreement becomes apparent when we compare the global features
of various correlation functions.
The difference between $\psi_{{\rm HM}}$ and the exact result is largest
for $J/t = 0$.  
These are in sharp contrast to $\psi_{{\rm BA}}$, with which
we can quantitatively reproduce both of the amplitude and 
the exponent of correlation functions.
This is because $\psi_{{\rm BA}}$ has the separation of charge 
and spin degrees of freedom correctly.
The effect of the spin-charge separation becomes clear especially 
in the small $J/t$ region.
In fact, when we compare the phase diagram determined by the present 
method with that of $\psi_{{\rm HM}}$,\cite{YO96}
the behavior of $K_\rho $ is much improved in the repulsive TLL region,
while it agrees with the result of $\psi_{{\rm HM}}$ 
in the attractive TLL phase.

Finally we mention some remaining issues.
   The spin wave function $Y( y_1,\dots, y_{M} )$ in Eq.~(\ref{psiBA})
has been approximated as a trial function of Jastrow-Marshall-type
for its simplicity.
It is known, however, a liquid state is realized in the 1D 
antiferromagnetic Heisenberg model.
 When a RVB-type trial state is used as the spin part 
in Eq.~(\ref{psiBA}) instead of Jastrow-Marshall-type, 
further improvement may be expected.
Actually, Ogata\cite{Ogata} has examined a trial function composed of 
a spin trial state of RVB-type and spinless fermions for $J/t=0$.
It is interesting to correlate this trial function with 
a Jastrow factor, applying to all range of $J/t$.

    An application of present method with some modifications 
to magnetic properties such as spin susceptibility and 
magnetization curve is also interesting in order to elucidate 
the metal-insulator transition in the 1D $t$-$J$ model.
In fact, it was shown that Jastrow wave functions reproduce charge 
and spin susceptibilities and magnetization curve correctly, 
in contrast with the Gutzwiller approximation.\cite{YO96}

An important question is whether the properties 
of strongly correlated electrons realized in 1D system
can be extended to higher dimensions
because of their close connection to high-$T_c$ 
superconductors.\cite{2DTLL}
In 2D system, it is not established even for the metallic regime
whether the ground state is the Fermi liquid or TLL.
In these contexts, an extension of the present method to 2D systems 
together with reexamination of the appropriate Hamiltonian
is under consideration.

   Quite recently, we have found that a part of our result for
the variational wave function of correlated singlet pairs 
has been obtained independently by Chen and Lee.\cite{Chen96}

\acknowledgments

One of the authors (K. K.) is grateful to H. Yokoyama 
for useful comments. 
A part of the numerical calculations was performed using the facilities
of the Computer Center, University of Tokyo.

%

%
%
%
\begin{figure}
\caption{
Optimization result of variational parameters 
in $\psi_{\protect{\rm BA}}$ is shown
for $n_e=\frac{1}{2}$ and $N_s=100$.
Triangles, squares, and circles represent $\nu_1$, $\nu_2$, and $\nu_s$,
respectively.
The solid lines are the least-squares fits of the data.
The transition to a phase-separated state is
shown by an arrow at $J/t \approx 3.3$.
}
\label{figopt}
\end{figure}

%
%
%
\begin{figure}
\caption{
(a) the momentum distribution function $n(k)$,
(b) the spin correlation function $S(k)$,
(c) the charge correlation function $C(k)$, and
(d) the singlet pairing correlation function $P(k)$ 
for $n_e=\frac{1}{2}$ and $ J/t = 0$.
Open circles and broken lines denote the VMC results of 
$\psi_{\protect{\rm BA}}$ and  $\psi_{\protect{\rm HM}}$, respectively, 
for $N_s=100$.
The exact results\protect\cite{OS,YO90,PS} 
for $N_s=52$ are also shown with solid lines.
}
\label{figcf0}
\end{figure}

%
%
%
\begin{figure}
\caption{
(a) the momentum distribution function $n(k)$,
(b) the spin correlation function $S(k)$,
(c) the charge correlation function $C(k)$, and
(d) the singlet pairing correlation function $P(k)$ 
for $n_e=\frac{1}{2}$ and $N_s=100$.
The data are evaluated by the VMC calculations 
with $\psi_{\protect{\rm BA}}$ for $J/t=1$ (open diamonds), 
2 (open squares), and 3 (open triangles).
The broken lines denote the corresponding results 
of $\psi_{\protect{\rm HM}}$.
}
\label{figcf123}
\end{figure}

%
%
%
\begin{figure}
\caption{
The VMC results of $ \ln |P(0)-P(k_1)|$ at $n_e=1/2$ are plotted
for $J/t=0$ (circles), 1 (diamonds), 2 (squares), and 3 (triangles)
as a function of $\ln N_s$, where $k_1=2\pi/N_s$.
The data are evaluated with $\psi_{\protect{\rm BA}}$ 
for $N_s=$ 12, 20, 36, 60, and 100.
The solid lines are the least-squares fits of the plots.
}
\label{figfp}
\end{figure}

%
%
%
\begin{figure}
\caption{
The phase diagram of the 1D $t$-$J$ model as determined by
$\psi_{\protect{\rm BA}}$ and $\psi_{\protect{\rm RVB}}$.
In the spin-gap phase,
$\psi_{\protect{\rm RVB}}$ is more stable than 
$\psi_{\protect{\rm BA}}$ while other phases are described 
by $\psi_{\protect{\rm BA}}$.
The curves represent the contours of constant correlation
exponent $K_\rho$ evaluated from Eq.~(\protect\ref{Kba}).
}
\label{figpd}
\end{figure}

%
%
%
\begin{figure}
\caption{
(a) the momentum distribution function $n(k)$,
(b) the spin correlation function $S(k)$,
(c) the charge correlation function $C(k)$, and
(d) the singlet pairing correlation function $P(k)$ 
for $n_e=0.2$ and $ N_s=150$.
Selected values of $J/t$ are 0 (triangles), 2.5 (squares), 
and 3 (circles) as typical cases of the repulsive TLL, 
attractive TLL, and spin-gap state, respectively.
}
\label{figcf02}
\end{figure}

%
%
%
\begin{figure}
\caption{
Exponents of the singlet pairing correlation function for $n_e=0.2$ 
as a function of $J/t$.
Squares are evaluated with $\psi_{\protect{\rm BA}}$,
while circles with $\psi_{\protect{\rm RVB}}$,
fitting $P(k)$ to Eq.~(\protect\ref{fitP})
for lattice sizes ranging from $N_s=30$ to 150.
Solid lines represent the value of $\nu_1+\nu_2+2$ for 
$\psi_{\protect{\rm BA}}$
and $ 2 \lambda -1$ for $\psi_{\protect{\rm RVB}}$,
where $\nu_1$,  $\nu_2$, and  $\lambda$ are the optimized 
variational parameters.
}
\label{figexp}
\end{figure}

%
%
%
\mediumtext
\begin{table}
\caption{
Ground-state energies of the 1D $t$-$J$ model 
in the limit $N_s \rightarrow \infty$ 
for the quarter-filled case ($n_e=\frac{1}{2}$).
The VMC results are obtained from the finite size 
scaling Eq.~(\protect\ref{fite}) with $N_s=$ 12, 20, 36, 60, and 100.
Exact results obtained from the Bethe ansatz 
for $J=0$\protect\cite{Shiba} and 2,\protect\cite{J2}
and extrapolated values of the exact diagonalization of 
small clusters\protect\cite{YO96} for $J=1$ and 3 
are listed as a comparison.
The unit of the energy is $t$.
}
\label{tableI}
\begin{tabular}
   {cdd@{${}\pm{}$}r@{}lr@{}ld@{${}\pm{}$}r@{}lr@{}l}
    $J$  &  \multicolumn{1}{c}{Exact}  &
    \multicolumn{3}{c}{VMC $E_{\infty} (\psi_{{\rm HM}})$}  &
    \multicolumn{2}{c}{(Error \%)} & 
    \multicolumn{3}{c}{VMC $E_{\infty} (\psi_{{\rm BA}})$}  &
    \multicolumn{2}{c}{(Error \%)}  \\  \tableline
    0   & $-$0.636620  &
        $-$0.6119  &   0&.0003   & 
        (4&)    & 
        $-$0.636620 &  0&&  
        (0&)   \\
    1   & $-$0.755359  & 
        $-$0.7493   &   0&.0001   & 
        (0&.8)  & 
        $-$0.75488  &   0&.00005 & 
        (0&.06) \\
    2   & $-$0.903649  &
        $-$0.90315  &   0&.00005  & 
        (0&.06) & 
        $-$0.90318  &   0&.00006 &
        (0&.05) \\
    3   & $-$1.081713  &
        $-$1.0774   &   0&.0001   & 
        (0&.4)  & 
        $-$1.0806   &   0&.0001  &
        (0&.1)  \\
\end{tabular}
\end{table}

\narrowtext
%
%
%
\begin{table}
\caption{
The correlation exponent $K_{\rho}$    
for $n_e=\frac{1}{2}$.
Results for $\psi_{\protect{\rm BA}}$ are evaluated from 
fitting $P(k)$ to Eq.~(\protect\ref{fitP})
for lattice sizes ranging from $N_s=12$ to 100.
With $\psi_{\protect{\rm HM}}$,
$K_{\rho}$  is directly obtained from Eq.~(\protect\ref{Khm}).
The last column is evaluated from the optimized 
variational parameters in $\psi_{\protect{\rm BA}}$.
}
\label{tableII}
\begin{tabular}{cddd@{${}\pm{}$}r@{}ld}
        &  \multicolumn{1}{c}{Expected} & 
           \multicolumn{5}{c}{} \\ 
  $J/t$ &  \multicolumn{1}{c}{value} & 
           \multicolumn{1}{c}{$K_{\rho}(\psi_{{\rm HM}})$} & 
           \multicolumn{3}{c}{$K_{\rho}(\psi_{{\rm BA}})$} & 
           \multicolumn{1}{c}{$(\nu_1+\nu_2 + 2)^{-1}$} \\ \tableline
    0  &  0.50\cite{Schulz,CFTJ0} & 0.40 & 0.51 & 0&.03 & 0.50 \\
    1  &  0.62\cite{OLSA}         & 0.57 & 0.62 & 0&.05 & 0.65 \\
    2  &  0.85\cite{CFTJ2}        & 0.95 & 0.85 & 0&.07 & 0.95 \\
    3  &  2.2\cite{OLSA}          & 3.2  & 2.2  & 0&.2  & 2.4  \\
\end{tabular}
\end{table}

%
%
%
\begin{table}
\caption{
The correlation exponent $K_{\rho}$ for $n_e=0.2$,
obtained by fitting $P(k)$ to Eq.~(\protect\ref{fitP})
for lattice sizes ranging from $N_s=30$ to 150.
The last column is evaluated from the optimized 
variational parameters in the trial wave functions.
}
\label{tableIII}
\begin{tabular}{dcd@{${}\pm{}$}r@{}ld}
  \multicolumn{5}{c}{} &
  \multicolumn{1}{c}{$(\nu_1+\nu_2 + 2)^{-1}$} \\
  \multicolumn{1}{c}{$J/t$} & Type of $\psi$ & 
  \multicolumn{3}{c}{$K_{\rho}$} &    
  \multicolumn{1}{c}{or $(2 \lambda )^{-1}$} \\
  \tableline
   0.0  & BA   & 0.51 &  0&.05 &  0.50    \\
   1.0  & BA   & 0.57 &  0&.07 &  0.63    \\
   2.0  & BA   & 0.97 &  0&.06 &  0.98    \\
   2.5  & BA   & 1.7  &  0&.2  &  1.6     \\
   2.8  & RVB  & 1.6  &  0&.2  &  1.7     \\
   3.0  & RVB  & 1.8  &  0&.2  &  1.9     \\
   3.2  & RVB  & 2.3  &  0&.4  &  2.2     \\
\end{tabular}
\end{table}

\end{document}